\documentclass[journal]{IEEEtran}
\usepackage{cite}
\usepackage{amsmath}
\usepackage{algorithm}
\usepackage{algpseudocode}
\usepackage{subfigure}
\usepackage{float}
\usepackage{graphicx}
\usepackage{booktabs}
\usepackage{epsfig} 
\usepackage{color}
\usepackage{url}
\usepackage{array}
\usepackage{caption}
\floatname{algorithm}{Protocol}
\begin{document}

\title{A Pvalue-guided Anomaly Detection Approach Combining Multiple Heterogeneous Log Parser Algorithms on IIoT Systems}%

\author{Xueshuo~Xie,\IEEEmembership{Member,~IEEE,}
	Zhi~Wang,\IEEEmembership{Member,~IEEE,}
	Xuhang~Xiao,\IEEEmembership{Member,~IEEE,}
	Lei~Yang,~\IEEEmembership{Member,~IEEE,}
	Shenwei~Huang,~\IEEEmembership{Member,~IEEE,}
	and~Tao~Li,~\IEEEmembership{Member,~IEEE}
	\thanks{X.Xie, Z. Wang, and T. Li are with the Department of Cyberspace Science, Nankai University, Tianjin, 300350, China. E-mail: zwang@mail.nankai.edu.cn.}
}



\maketitle

\begin{abstract}

	Industrial Internet of Things (IIoT) is becoming an attack target of advanced persistent threat (APT). Currently, IIoT logs have not been effectively used for anomaly detection. In this paper, we use blockchain to prevent logs from being tampered with and propose a pvalue-guided anomaly detection approach. This approach uses statistical pvalues to combine multiple heterogeneous log parser algorithms. The weighted edit distance is selected as a score function to calculate the non-conformity score between a log and a predefined event. The pvalue is calculated based on the non-conformity scores which indicate how well a log matches an event. This approach is tested on a large number of real-world HDFS logs and IIoT logs. The experiment results show that abnormal events could be effectively recognized by our pvalue-guided approach.
	
\end{abstract}

\begin{IEEEkeywords}
	anomaly detection; conformal prediction; log parser; IIoT security
\end{IEEEkeywords}

\section{Introduction}

IIoT systems  are becoming the attack targets of advanced persistent threat (APT) that results in the  damage of equipment operations and the threat to information security. Due to the fact that IIoT devices are usually isolated from the Internet, network security problems of IIOT have not been paying enough attention, especially the anomaly detection of system logs.


System logs record detailed system states and events, which can be used to identify system anomalies, trace system behaviors, troubleshoot failures and performance issues. In order to hide malicious trails, attackers have to disguise their behaviors in log records in target system. Traditionally, logs in IIoT systems are separately saved on different devices. The log storage approach brings two challenges for log security analysis in IIoT systems: (1) logs are easy to tamper with; (2) the lack of effective log sharing and analysis strategy hinders log analysis from a holistic perspective, limiting the power of multiple log analysis algorithms.


Blockchain provides a reliable solution for consistent, distributed data storage in an untrusted environment. It could effectively combat log tampering attacks~\cite{bahga2016blockchain, teslya2017blockchain}. Usually, logs generated by different devices are unstructured and should be preprocessed into standard raw logs. The use of blockchain could prevent logs from being tampered by attackers and enable log sharing between different IIoT devices.


Generally, different IIoT devices and applications generate log message of different data structures, including various format of labels (message, warning, etc.), timestamps, ID and unstructured message content. Log parser algorithms can extract log event templates from unstructured message content without the source code of systems and applications. Log event templates can be used for anomaly detection in many areas of research. Currently, there are many log parser algorithms, such as  LKE~\cite{fu2009execution}, IPLoM~\cite{Makanju2012A}, LogSig~\cite{tang2011logsig}, Drain~\cite{he2017drain}, DrainV1~\cite{vaarandi2015logcluster} etc. All log parser algorithms target raw log messages that consist of two parts including a constant part and a variable part. The constant part of a log message is used to form an event template. In [16], the authors give an evaluation study on log parsing algorithms in accuracy and efficiency. Each algorithm can effectively process log messages stored in blockchain, to obtain diverse event template set.


There are already many machine learning algorithms used for log anomaly detection, such as LR~\cite{bodik2010fingerprinting}, Decision Tree~\cite{chen2004failure}, SVM~\cite{liang2007failure},  Isolation Forest~\cite{liu2008isolation}, PCA~\cite{xu2009largescale}, Invariants Mining~\cite{Lou2010Mining}, Clustering~\cite{lin2016log}, DeepLog~\cite{Min2017DeepLog}, AutoEncoder~\cite{Borghesi2018Anomaly}, etc. In\cite{he2016experience}, the authors give an evaluation study on various anomaly detection models. Existing models are mainly used for anomaly detection of large-scaled distributed systems with huge log data and complex system scale. However, differed from distributed systems, IIoT has the following characteristics: (1) IIoT  runs continuously for most of the time and usually abnormalities occur very rarely; (2) the amount of log data generated by IIoT is also very small compared to large-scale distributed systems. However, IIoT abnormalities usually result in very serious security incidents.


We propose a pvalue-guided hybrid model for multiple heterogeneous log parser algorithms. To prevent log tampering and share logs among different devices, we introduce blockchain to store raw logs. The combination of multiple heterogeneous log parser algorithms effectively improves the detection accuracy. We have obtained log data from a real oil industry in a period of time. Experiments show that our could effectively detect  abnormal logs.

In summary, the main contributions of this paper are the following:
\begin{itemize}
	\item We design a pvalue-guided hybrid approach combining multiple log parser algorithms, which can effectively detect abnormal logs of a real oil industry and a HDFS system. 
	\item The string weighted edit distance is selected as a measure to compute the non-conformal score of a log to an event. And we use conformal prediction to calculate pvalues for anomaly detection.
	\item We design and implement a prototype system for IIoT anomaly detection, and test the performance on HDFS and IIoT logs.
\end{itemize}


The remainder of this paper is organized as follows: Section 2 gives a brief review on log parsers, anomaly detection, conformal prediction and blockchain. Section 3 describes the overview of the model. Section 4 gives a detailed description of the pvalue-guided approach, including log preprocessing, conformal prediction, statistical analysis and anomaly detection. We present the experiment results in Section 5. Finally, Section 6 concludes the paper.

\section{Related work}

Log analysis, consisting of log collection, log parsing and log mining \cite{he2017end}, can effectively identify system problems \cite{beschastnikh2011leveraging, shang2013assisting}. 

\textbf{Log parser algorithms}. As a key prerequisite for log mining, log parser algorithms are often used to extract the constant part of raw logs and form an event template set. Each algorithm uses different principles for the extraction of log templates. The most used algorithms include frequent pattern mining~\cite{nagappan2010abstracting} and LogCluster\cite{vaarandi2015logcluster}), and clustering~\cite{fu2009execution, tang2011logsig, hamooni2016logmine, Makanju2012A, he2017drain, he2018directed, vaarandi2015logcluster}. A new log parser algorithm MoLFI [40] uses the evolutionary algorithms and achieves a good result. 


The log parser algorithms have been widely studied in recent years, but there still lacks evaluation methods for multiple algorithm cooperation. In\cite{he2016evaluation}, the authors give four log parser algorithms (SLCT, IPLoM, LKE and LogSig) an evaluation on their effectiveness and accuracy. Their experiment codes and log datasets are publicly available. In \cite{zhu2018tools} , they have made a further research on log parser algorithms evaluation of 13 log parser algorithms on a total of 16 log datasets.



\textbf{Anomaly Detection}. As an important branch of log analysis, anomaly detection is widely used in large-scaled distributed systems. It has also formed a complete analysis process including log collection, log parsing, feature extraction and machine learning. Based on different design techniques, anomaly detection models are mainly divided into supervised models and unsupervised models. In order to select an efficient and accurate anomaly detection in practical applications, \cite{he2016experience} provide a detailed review and evaluation of 6 state-of-art log anomaly detection methods. 


\textbf{Conformal prediction}. As described in~\cite{shafer2008tutorial}, the authors propose the theory of conformal prediction that uses past experience to determine precise level of confidence in new prediction. Conformal prediction uses a non-conformal measure to calculate the non-conformity score. We can obtain a pvalue that can be used to make decisions or evaluation, whose outputs is a prediction set with fixed confidence level. In~\cite{fedorova2012plug}, the authors use conformal prediction to detect deviation of data sequence under the assumption that data are independent and identically distributed. Conformal evaluation uses the theory of conformal prediction to measure the non-conformity of a test object to a class compared to all other objects. By using conformal evaluation, the authors detect concept drift and identify aging classification in malware classification models~\cite{jordaney2017transcend}. Conformal evaluation provides better understanding for model quality.



\textbf{Blockchain in IIoT}. Recently, blockchain has attracted attentions for its accomplishment in cryptocurrencies and distributed applications. There are many studies on the combination of blockchain and IoT to provide a reliable solution in IIoT systems. In\cite{afanasev2018application}, the researchers  use smart contracts in an industrial production audit system. As consortium blockchain uses many security mechanisms such as identity and member service provide, it can be used for secure energy trading in industrial internet of things.

\section{Model overview}

\begin{figure*}[ht]
	\includegraphics[width=.8\textwidth]{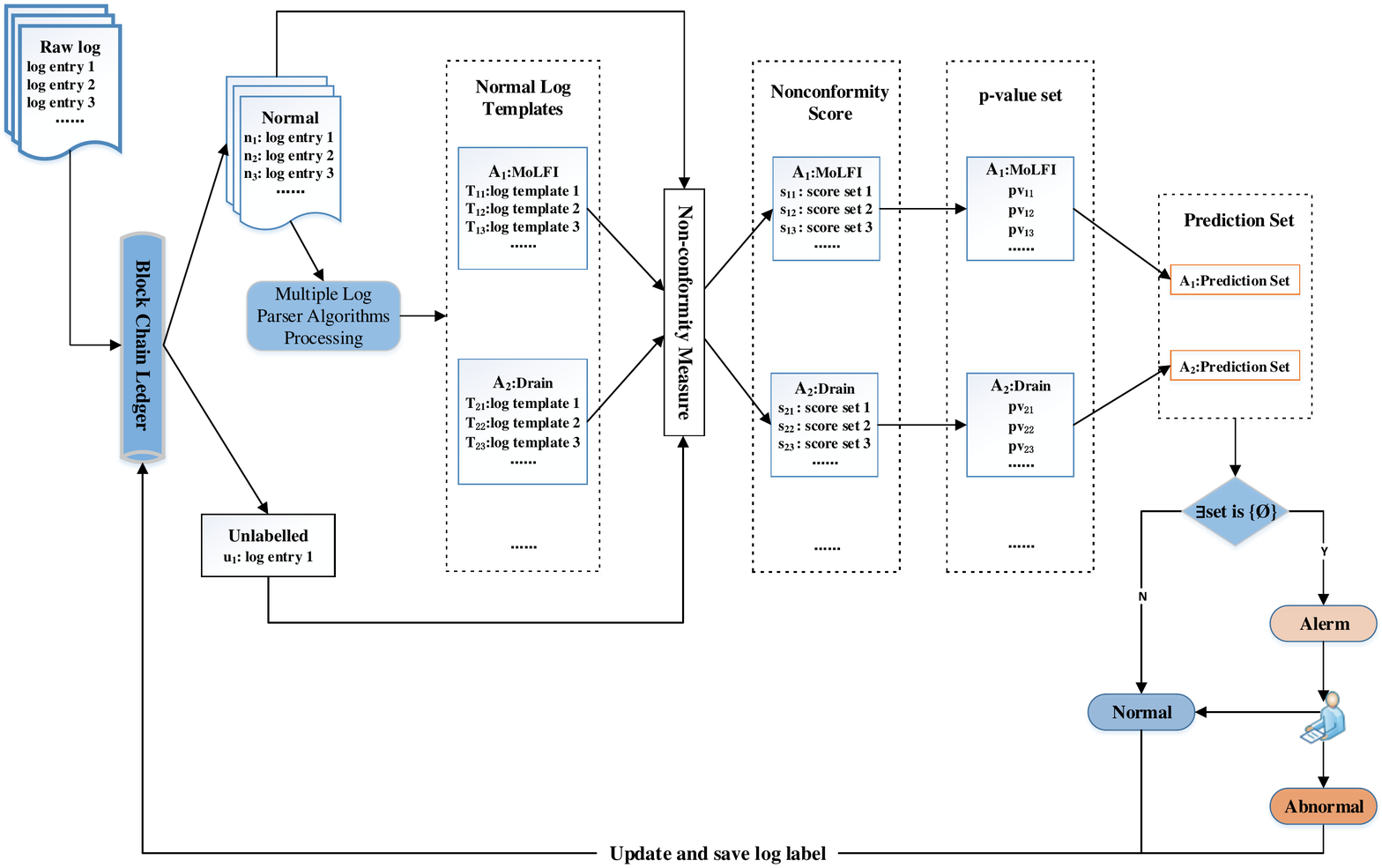}
	\caption{The core of the pvalue-guided approach includes log parser, non-conformal measure and pvalue-guided anomaly detection.}
	\label{fig:Model}       
\end{figure*}

Fig.1 gives the overall of the pvalue-guided approach for multiple heterogeneous log parser algorithm. The core of the model mainly includes event template generation, the design of non-conformal measure and anomaly detection based on pvalues.


\textbf{Logs stored in blockchain}. IIoT system consists of various devices that routinely generate log messages to record information of applications and systems. Different devices and applications may generate log messages of different structures. These unstructured log messages are preprocessed into standard raw logs by automated scripts. They are stored in the blockchain according to certain rules.



\textbf{Heterogeneous log parsers}. Log parser algorithms can extract a set of log event templates from raw logs. we use multiple heterogeneous log parser algorithms to  train normal logs to get an event template set. The event template sets generated by different log parsers are different. 
And the logs produced by systems tend to grow in numbers with diverse event over time, which will lead to the model aging problem. By combining multiple heterogeneous log parsers, the quality of training results could  be improved at accuracy without analysis on the source code of the system and application.

\begin{figure*}[htp]
	\includegraphics[width=1\textwidth]{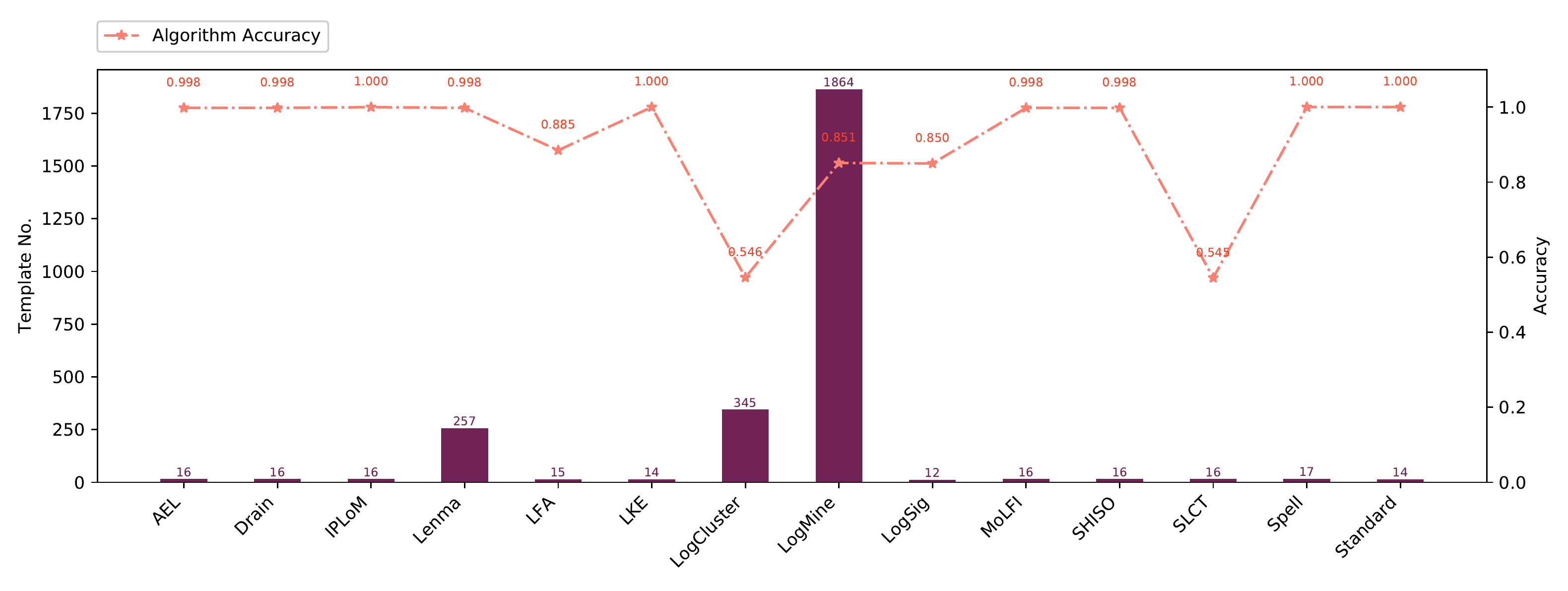}
	\caption{The 13 log parser algorithms used on the HDFS\_2K logs generate the template number and algorithm's accuracy. As the algorithm’s principle is different, the generated event templates is different on quantity and content.}
	\label{fig:Template}       
\end{figure*}



\textbf{Conformal prediction} . The core of conformal prediction is a non-conformal measure that calculates the non-conformal score between a log  and an event template. In our paper, we choose the weighted edit distance as the non-conformity measure — a real-valued scoring function ${A_D}( {C,z^ *} )$. We can obtain a  pvalue that can be used to make statistical decisions and give a prediction set  of the fixed confidence level $1 - \varepsilon$. $\varepsilon$ is the significant level selected by user which indicates the maximum probability of error. Conformal prediction~\cite{shafer2008tutorial} could give new  logs a series of pvalue set based on known event templates. The number of pvalue sets is the same as the number of selected log parser algorithms. The number of the set element is the same as the number of templates generated by the log parser.


We apply the significant level $\varepsilon$ on the pvalue set to filter pvalues lower than $\varepsilon$ from the set. If the pvalue set is not empty, the new log is a normal log; if the pvalue set is empty, an alarm is generated. Then human analysis is needed to judge whether the log is a real abnormal event.

\section{Methodology}


In this section, we give a detailed description of our approach including log preprocessing, heterogeneous log parsers, conformal prediction and pvalue-guided anomaly detection. The weighted edit distance is selected as the non-conformity measure to calculate pvalue. Finally, we use the pvalue sets for anomaly detection based on a given significant level.

\subsection{Log preprocessing}

In IIoT system, there are many devices that  generate a large amount of log records. Different devices and different applications generate log messages in different formats. But there are some common parts, such as timestamp, unique ID (user or machine) and  unstructured message content. Therefore, before log storage in the blockchain, a preprocessing is required to generate a raw log in standard format. These logs can be used to discover and identify system anomalies, trace system behaviors and malicious attacks.

\subsection{Heterogeneous log parsers}

Log parser is an important part of our anomaly detection approach. Heterogeneous log parsers could analysis logs from different perspectives. As a key prerequisite for complex log mining, log parser algorithms are used to extract the constant part of a raw log and generate an event template set. There are many algorithms (SLCT, POP, LKE, LogSig, IPLoM, Drain, etc) used in different scenarios. 

Instead of obtaining system source code for analysis, log parser only generate event templates by analyzing existing log messages.  To improve the  accuracy of anomaly detection, we need to design a method to evaluate and combine the results given by multiple heterogeneous log parsers. In this paper, we select four heterogeneous log parsers (IPLoM~\cite{Makanju2012A}, LogCluster~\cite{vaarandi2015logcluster}, AEL~\cite{jiang2008abstracting, jiang2008automated} and Spell~\cite{Min2017Spell}) to test our approach. 


\subsection{Conformal prediction}
In this part, we  combine multiple log parser algorithms to improve the accuracy of anomaly detection result. We introduce conformal prediction \cite{shafer2008tutorial} method to analysis the quality of log parser for a unknown log. Firstly, we choose a non-conformity measure, weighted edit distance, to measure the conformity between a log message and known event templates. We use conformal prediction to calculate a pvalue set which indicate the quality of each log parser result from the statistical perspective.


Non-conformity measure is a real-valued function used to calculate the conformity between a group of messages belonging to the same class and a new log message. Given a new log message l, $A_D(C,l^*)$  outputs a non-conformal score, where $D$ is the training logs and $T$ is the template set generated by log parsers. The scoring function is denoted as:

$$\alpha_{l^*} = A_D(T,l^*) \eqno(1)$$


When a new log message is generated we calculate the non-conformity score between templates and new log. For two string sequences l1 and l2, the score is denoted as:
$$Score({l_1}, {l_2}) = \sum\limits_{i = 1}^n {\frac{1}{{1 + {e^{({x_i} - v)}}}}}\eqno(2)$$


Where n is the number of necessary operations (add, delete and replace) between  $l_{1}$ and $l_{2}$, $x_i$ is the index of the word that is operated by the  $i^{th}$  operation $n$, $v$ is a parameter controlling weight function (related to the length of log messages and templates).



We calculate the non-conformity score set using equation (2). For the training log messages, we obtain multiple non-conformity score set whose size in equal to the number of log parser algorithms. For the detection log message, we also  obtain the non-conformity score between the templates generated by multiple log parser algorithms. For a set of training logs $K$, the pvalue $p_{l^*}^{T}$ for a new log $l^*$ is the proportion of logs in class $K$ that are at least as dissimilar to other logs in $T$ as $l^*$. The computation of pvalue for the new log is denoted as:

$$\forall i \in K, {\alpha _i} = {A_D}{\rm{ }}\left( {T\backslash {l_i},{l_i}{\rm{ }}} \right)\eqno(3)$$
$${p_{l^*}^{T}} = \frac{{\# \{ i = 1,...,n|{\alpha _i} \ge {\alpha _{l^*}}\} }}{|K|}\eqno(4)$$

With the input of a new log, we obtain a set of non-conformal scores between the template set and the new log. We also calculate a pvalue set by equation (4), which can be used for anomaly detection based on statistical analysis. 

\begin{table*}[htp]
	\caption{The experimental dataset used to evaluate our model.}
	\centering
	\begin{tabular}{ccccc}
		\toprule
		\textbf{Dataset}	& \textbf{Description}	& \textbf{\#Messages}	& \textbf{\#Training messages}    &\textbf{\#Detection messages}\\
		\midrule
		HDFS		&Hadoop distributed file system log			&2000			&2000		&2\\
		IIoT		&real oil industrial system log			&132602			&100000		&32602\\
		IIoT small		&real oil industrial system log			&11000			&10000		&1000\\
		\bottomrule
	\end{tabular}
\end{table*}

\subsection{Anomaly detection based on pvalues}


By using conformal prediction, we obtain a multiple pvalue sets, the number of which equal to the template number generated by all chosen log parsers. When we apply the anomaly analysis of the pvalue set, we set an appreciate significant level to filter the pavlues lower that the significant level from pvalue sets. The prediction set consists of the template label which can be used to give a new log label. This includes two cases: (1) if the new log is normal and the pvalue is higher than the significant level, the prediction set contains the template label given by the log parser algorithm; (2) if the log is abnormal, its pvalues are all lower than the significant level and the prediction set is empty.

\begin{figure*}[htp]
	\centering
	\includegraphics[width=.8\textwidth]{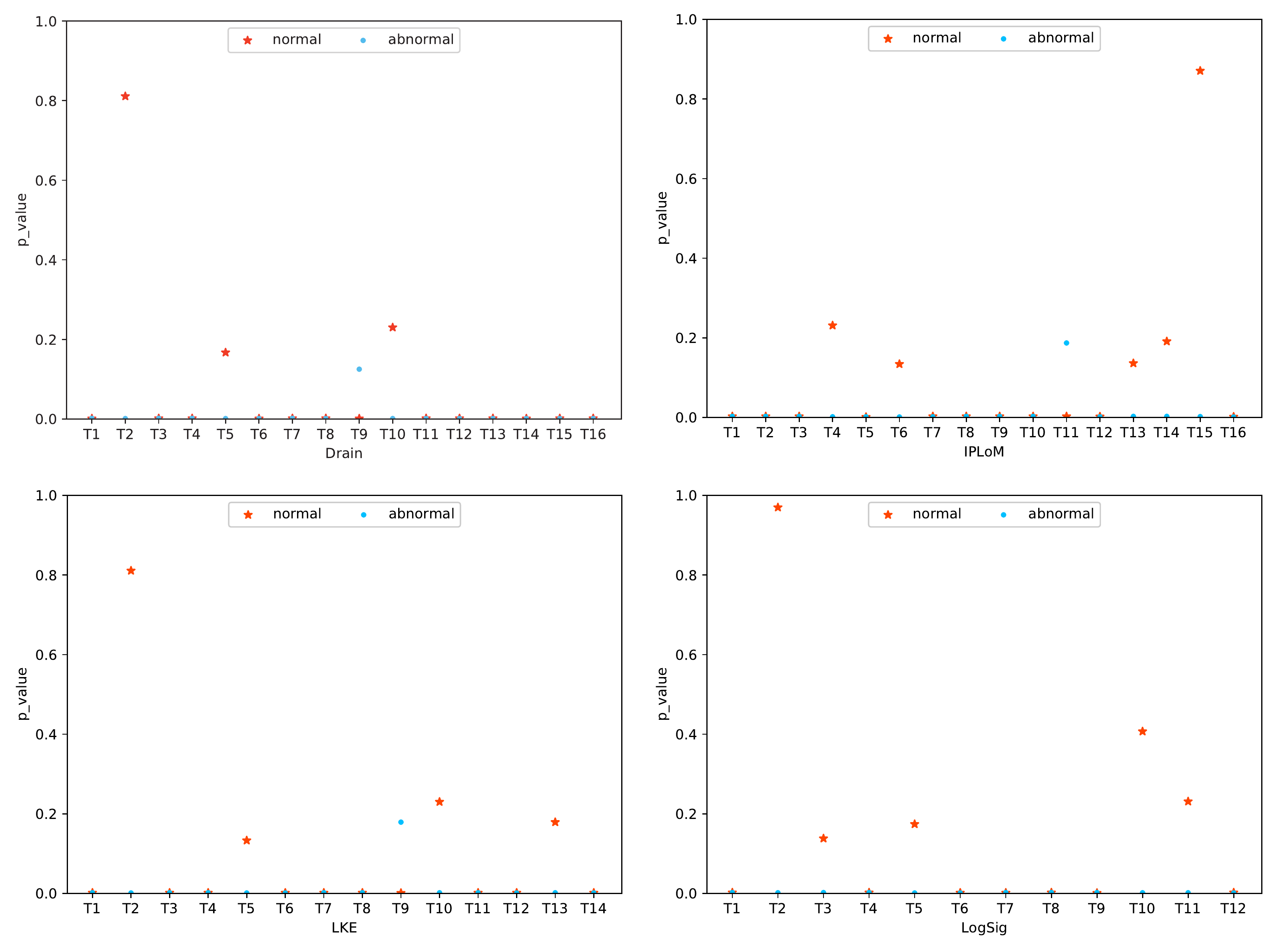}
	\caption{The p\_values calculate by four log parser algorithms. For the normal log message, each algorithm gives a high p-values of the match template; but for the abnormal log message, all algorithms gives a low p-values of each template in template set.}
	\label{merge}       
\end{figure*}

\section{Evaluation}

In this section, we evaluate our anomaly detection approach using both precision and recall. We apply this approach on a public HDFS log dataset and a dataset of  real oil industrial system logs.

\subsection{Experimental data}

Currently, there is no public IIoT log dataset. For the experiment, we got an IIoT log dataset from an oil company. And we also choose the public HDFS data as a benchmark for our experiment. The HDFS dataset contains 11,175,629 logs. Using log parser algorithms, we obtain a set of event templates. Using our anomaly detection model, we  calculate the pvalue of  new logs under different log parsers. For normal log messages, our approach will give  a higher pvalue. 
As Table I shows, we obtained 132,602 real oil industry system logs. It consists of  timestamp,  system ID,  Ethernet port ID, and an unstructured content recording the detailed operations. In the logs, we found two major anomaly logs: file errors (”Load *.ini failed!”) and data complete check errors (”check data complete failed!”. The file error is contained in one log and the data complete check error is contained in 787 logs. Since the computational non-conformity score (weighted edit distance) is time-consuming, we selected a small data set to test whether our model can effectively detect the 788 anomaly logs. In the small test data set, we used 10,000 normal log records as the training set, and 1000 logs containing 788 abnormal log records as the test set.

\subsection{The anomaly detection case on HDFS\_2k}

In this part, we mainly test the accuracy of our approach using a public HDFS dataset, that is, whether pvalue can accurately reflect the matching of log messages and log templates. If a  log message satisfies a template, the log will get a high pvalue from conformal analysis, otherwise it will get a low p value. we choose a normal log message and construct an abnormal log message (”*:Throw error while serving blk * from *”). we use our anomaly detection model to calculate the pvalues between each element of all template sets and the new log message. 
As Fig.~\ref{merge} shows, all log parser algorithms give a high pvalue to the normal log and its event template. For the abnormal log, all pvalues to the event templates are very low. It shows that using pvalue, logs can be classified to the correct event template. The normal log matches LogSig’s template T2 with pvalue 0.97, LKE’s template T2 with pvalue 0.81, IPLoM’s template T15 with pvalue 0.87 and Drain’s template T2 with pvalue 0.81. It also shows that using pvalue can help us find abnormal log records. LogSig, LKE, IPLoM and Drain all give low pvalues to abnormal log message, which is much lower than 0.2. This means that the abnormal log does not match any known event template. Through the test results on the HDFS dataset, we could find that pvalue-guided approach could effectively distinguish normal logs and abnormal logs.

\subsection{The anomaly detection case studies on IIoT logs}

In this part, we show the results of our approach on the real oil industry logs, which can effectively detect abnormalities such as file errors and data complete check errors. We compare the results of our multiple heterogeneous log parsers with the results of just using a single log parser. The experiment results show that higher precision and recall can still be achieved with the higher significance of our model settings.

\begin{table}[h]
	\caption{The Recall Rate of our model compared with the single log parser algorithm's result.}
	\centering
	\begin{tabular}{cccccc}
		\toprule
		\textbf{Significance}	& \textbf{0.27}	& \textbf{0.3} 	& \textbf{0.4}    &\textbf{0.6}	& \textbf{0.83}\\
		\midrule
		IPLoM		&0.9987			&-			&-		&-		&0.9987\\
		Spell		&-			&-			&0.9987		&1		&-\\
		AEL		&-			&0.9987			&-		&-		&-\\
		LogCluster		&-			&0.9987			&-		&-		&-\\
		Our Model		&0.9987			&0.9987			&1		&1		&1\\
		\bottomrule
	\end{tabular}
\end{table}


As Table II shows, our model can achieve a higher when we set the significance larger than 0.4 compared to the single log parser algorithms. Our model can set a wide range of significance. For the single log parser algorithms to do anomaly detection, there are only some significance points that  can be set for effectively detection. for our model, we can choose an appreciate significance for statistical analysis to get a larger. Our model can detect more anomalies. By setting a reasonable significance for statistical analysis, our model can detect 788 abnormal logs, and a single log detection algorithm often some abnormal log records for abnormal detection.

\begin{table}[h]
	\caption{The Precision Rate of our model compared with the single log parser algorithm's result.}
	\centering
	\begin{tabular}{cccccc}
		\toprule
		\textbf{Significance}	& \textbf{0.27}	& \textbf{0.3} 	& \textbf{0.4}    &\textbf{0.6}	& \textbf{0.83}\\
		\midrule
		IPLoM		&0.9862			&-			&-		&-		&1\\
		Spell		&-			&-			&0.9776		&0.9656		&-\\
		AEL		&-			&0.9813			&-		&-		&-\\
		LogCluster		&-			&0.9447			&-		&-		&-\\
		Our Model		&0.9704			&0.9259			&0.9195		&0.8955		&0.8347\\
		\bottomrule
	\end{tabular}
\end{table}



As Table III shows, our model can also achieve a higher precision when compared to the single log parser. Although the precision of our approa has a certain decline when set a large significance, the recall rate is increasing. Even if a single log parser algorithm has a higher precision for anomaly detection,  some of the abnormal logs will be misclassified. Such false negative usually is a very serious security incident in IIoT system. Because IIoT has a very high security requirement.

In summary, our anomaly detection approach can effectively detect the anomaly logs (file errors (”Load *.ini failed!”) and data complate check errors (”check data complate failed!”)) in IIoT systems.

\section{Conclusion}


Log records play an important role in the IIoT anomaly detection. However, traditional log records are easy to tamper with and not conducive to collaborative defense between different devices. To mitigate this problem, we propose a novel approach for IIoT anomaly detection which introduces the conformal prediction. The new system combine multiple log parsing algorithms based on pvalues in the conformal prediction. The approach was tested on the public HDFS log dataset and an IIoT log dataset, and the results show that abnormal events could be effectively recognized.

\end{document}